\documentclass[preprint2]{aastex}

\slugcomment{N}
\shorttitle{Trans vs Reflec}
\shortauthors{Pall\'e et al., 2005}

\usepackage{lscape}
\usepackage{natbib}

\begin{document}

\title{Characterizing the atmospheres of transiting rocky planets around late type dwarfs}

\author{E. Pall\'e\altaffilmark{1,2}, M.R. Zapatero Osorio\altaffilmark{3},  A. Garc\'ia Mu\~noz\altaffilmark{1,2}}

\affil{(1) Instituto de Astrof\'isica de Canarias, La Laguna, E38205,
Spain. }
\affil{(2) Departamento de Astrof\'isica, Universidad de La Laguna. Av. Astrof\'isico Francisco S\'anchez,
s/n E38206 -- La Laguna, Spain}\email{epalle@iac.es, agm@iac.es}
\affil{(3) Centro de Astrobiolog\'ia, CSIC-INTA, Madrid,
Spain. }\email{mosorio@cab.inta-csic.es}

\begin{abstract}

Visible and near-infrared spectra of transiting Hot Jupiter planets have recently been observed, revealing some of the atmospheric constituents of their atmospheres. In the near future, it is probable that primary and secondary eclipse observations of Earth-like rocky planets will also be achieved. The characterization of the Earth's transmission spectrum has shown that both major and trace atmospheric constituents may present strong absorption features, including important bio-markers such as water, oxygen and methane. Our simulations using a recently published empirical Earth's transmission spectrum, and the stellar spectra for a variety of stellar types, indicate that the new generation of extremely large telescopes, such as the proposed 42-meter European Extremely Large Telescope(E-ELT), could be capable of retrieving the transmission spectrum of an Earth-like planet around very cool stars and brown dwarfs ($T_{\rm eff} \le \sim 3100$\,K). For a twin of Earth around a star with $T_{\rm eff}\sim 3100 K$ (M4), for example, the spectral features of $H_2O$, $CH_4$, $CO_2$, and $O_2$ in the wavelength range between 0.9 and 2.4 $\mu m$ can simultaneously be detected within a hundred hours of observing time, or even less for a late-M star. Such detection would constitute a proof for the existence of life in that planet. The detection time can be reduced to a few hours for a super-Earth type of planet with twice the Earth's radius.

\end{abstract} 

\section{Introduction}

Since the discovery in 1992 of the first planets outside the solar system (Wolszczan and Frail 1992), the number of planet detections has steadily increased. Although we are not capable of detecting and exploring planets like our own yet, ambitious missions, both ground- and space-based, are already being planned for the next decades, and the discovery of Earth-like planets is probably just a matter of years. Once these planets are found, the efforts will concentrate on the detection and characterization of their atmospheres, including the possibility of finding conditions suitable for life. This will be achieved by either direct detection of the light reflected/emitted by the planet or, if the planet has transits, by in- and out-of-transit comparative spectroscopy (Brown, 2001).

%The first extrasolar giant planet detected by direct observation was announced by Chauvin et al. (2004). In 2008, two planetary systems were directly imaged with different instrumentation and almost simultaneously (Kalas et al., 2008; Marois et al. 2008). Nevertheless, direct detection is an extremely challenging task. At present, the technological development is not sufficient to resolve planets that lie close to the parent star. Thus, so far, the only technologically practical tool for exoplanet atmospheric characterization is the exploitation of the special geometry of transiting planets.

Struve (1952) first proposed to search for extrasolar planets using the method of transits, but it has been only recently that several planets, all far larger than the Earth, have been partly characterized using this technique (Charbonneau et al., 2002; Swain et al, 2008 and many more). The technique consists in making precise observations when the planet passes in front of the star (primary eclipse or transit), or before and after the planet is hidden behind the star (secondary eclipse).

Observations of primary transits allow us to obtain the transmission spectrum of a planet atmosphere. When the transit occurs part of the light that reaches the observer has crossed the planetary atmosphere and contains the signatures of its spectroscopically-active component gaseous species. Based on HST high-precision spectrophotometric observations, Charbonneau et al. (2002) detected the absorption from sodium in the atmosphere of HD209458b, while Richardson et al (2007) reported its infrared spectra (7.5-13.2 $\mu m$). Using the Spitzer telescope and this same methodology, Tinetti et al (2007) and Swain et al (2008) have recently reported the presence of water and methane, respectively, in the atmosphere of planet HD189733b, first discovered by Bouchy et al. (2005).

Important information about the planet's albedo and large-scale meteorology can also be derived from secondary eclipses, when the planet passes behind the parent star. Observations just before and during the eclipse have allowed for the characterization of the reflection/emission spectrum of the planet itself. Harrington et al. (2007), for example, reported the direct detection of thermal emission from HD149026b, the hottest exoplanet discovered so far ($2,300 \pm 200 K$).

The transmission and emission spectra of exoplanets offer complementary information to understand the composition, thermal balance, and dynamics of their atmosphere. One of the best studied hot Jupiter planets is HD189733b. Swain et al. (2008) reported the first near-infrared transmission spectrum of that planet, which showed the presence of methane and water. Although on thermochemical grounds, carbon monoxide was expected to be abundant in its upper atmosphere it was not identifiable in the spectra. One year later, Swain et al. (2009), again observed the dayside spectrum of HD189733b during a secondary eclipse.  They detected the presence of  water ($H_2O$), carbon monoxide ($CO$), and carbon dioxide ($CO_2$), but the signature of methane ($CH_4$) was not distinguishable.

But what about the detection and characterization of small, rocky planets? There are presently two space mission capable of finding transiting rocky planets: $CoRoT$ and $KEPLER$. The latter is expected to retrieve between 50 and 640 terrestrial inner-orbit planets, depending on the as-of-yet unknown mass and radius distribution of these planets (Borucki et al, 2009). While most of the targets will be around G and K stars, if a planet was confirmed around an M star, it would immediately become a candidate for follow-up atmospheric characterization because of the smaller star/planet contrast ratio. Plato, another space-based mission searching for transiting planets (Catala, 2009), could also find suitable targets, but the mission is still pending approval. Finally, ground-based searches for transiting rocky planets can also provide confirmation of suitable candidates. Transits of planets of size about twice the Earth's radii have already been observed (Leger et al, 2009; Charbonneau et al, 2009). In particular, the MEarth project (Charbonneau et al, 2009) has detected a transiting super-Earth planet around an M3-type star.

%In the foreseeable future, the James Webb Space Telescope (JWST) will be another great tool to characterize extrasolar planets through transit spectroscopy. However, although the JWST would be able to do a rough characterization of transiting super-Earths around M stars, this is only true for the closest M stars (within 10-20 pc) (REFS).

\begin{figure*}[h]
\epsscale{2.0}  \plotone{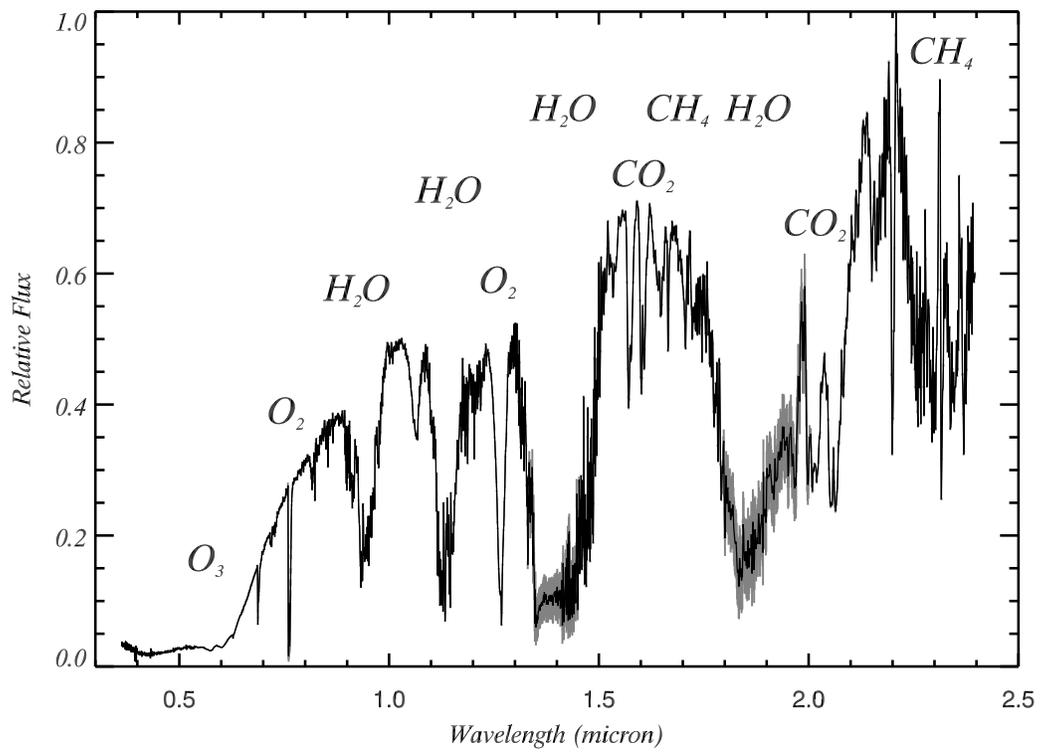}
\caption{Earth's visible and near-infrared transmission spectra. The Earth's transmission spectrum is a proxy for Earth
observations during a primary transit as seen beyond the Solar System. The absorption features of some of the major atmospheric constituents are marked. Adapted from Pall\'e et al. (2009)}\label{fig1}
\end{figure*}

Recently, the optical and near-infrared transmission spectrum of the Earth, was obtained during observations of a lunar eclipse (Pall\'e et al, 2009). It was found that, during the transit, some of the Earth's weakest atmospheric features become much more prominent than in the reflected spectrum. Atmospheric molecules such as $O_3$, $O_2$, $H_2O$, $CO_2$, and $CH_4$, are readily detectable (see Figure~\ref{fig1}). Thus, in the optical and near-infrared spectral ranges, the transmission spectra of the Earth ($T_{\rm eff} \sim255 K$) yields much more information content with regards to atmospheric characterization that its reflectance spectrum.

Up to now, observations have revealed a wealth of planetary systems, and a wide variety in the physical properties of giant planets. A variety of rocky planet atmospheric composition and chemistry is also expected to be found when their characterization becomes feasible. In particular, for a rocky planet around an M star, since the chemistry of species like $O_3$ and $CH_4$ involves photochemistry and the high probability of the system being tidally locked, its atmospheric composition and dynamics could be quite different than in the Earth's case (Segura et al, 2005; Tarter et al, 2007). However, in order to restrict the number of possible scenarios we have restricted ourselves to an Earth's twin case. Here, we use a semi-empirical approach to explore the feasibility of  characterizing, with current and planned telescopes facilities, the atmospheres of such earth-like planets beyond our solar system.

\section{Detection and characterization of rocky planets in the near future}

In order to characterize a planetary atmosphere through direct detection of the light either emitted by or reflected from the planet, it becomes necessary to remove the large stellar contribution to the photon flux. In the best case scenario, supposing a fully illuminated Earth disk, the contrast ratio between the Sun and the Earth is of the order of  $10^{-10}$.  This number is obtained following,

%\begin{equation}
%\frac{F_E}{F_S} = \frac{S a Ae}{L} = \frac{L}{4 \pi d_{S-E}^2}  \frac{a A_E}{L}, \label{eq1}
%\end{equation}

\begin{equation}
\frac{F_e}{F_s} = a \left( \frac{ R_{e}}{d}\right)^{2} +  \frac{B(T_e)}{B(T_s)} \left( \frac{R_e}{R_s}\right)^2, \label{eq1}
\end{equation}

\noindent where $R_e$ is the Earth's radius, $R_s$ is the Sun's radius,  $d$ is the Sun-Earth distance, and $a$ is the earth's geometric albedo (Qiu et al, 2003) and $B(T_e)$ and $B(T_s)$ are the intrinsic emission for the Earth and the Sun, respectively (blackbody-like emission in the case of the Earth). In the visible and near-infrared wavelength ranges discussed in this paper, the second term of the equation is $\le \le 10^{-15}$; for the Earth, this contribution is negligible as compared to the reflected, or even transmitted, stellar flux. At mid-infrared wavelengths, where the intrinsic luminosity of the  Earth is significantly larger, the situation may be different (Tinetti et al, 2006). Thus from hereon, the thermal emission from the planet is neglected from our formulation and discussions. With only a fraction of the illuminated planetary disk visible, the contrast ratios may be closer to $10^{-11}$.

If the same planet is observed during a transit, according to the area ratios between the Sun and the solid core of Earth, there will be a solid decrease in stellar light, $\Delta F_{sc}$, of the order of $10^{-5}$

\begin{equation}
\Delta F_{sc} = \frac{A_{Ec}}{A_S} = \left(\frac{ R_{e}}{ R_{s}}\right)^2, \label{eq2}
\end{equation}

If the planet has an atmosphere, the atmosphere will also block some of the stellar light. This decrease can be calculated as:

\begin{equation}
\Delta F_{a} = \frac{ (R_{e} + h)^2 - R_{e}^2}{ R_{s}^2}, \label{eq3}
\end{equation}

\noindent where $h$ is vertical extent of the planet's atmosphere. Supposing an atmosphere extending to 40 $km$ height,  the area ratio, $\Delta F_{a}$,  is of the order of $1.05$ $10^{-6}$ (Ehrenreich et al., 2006). For saturated atmospheric features, such as some water bands on Earth, the transit would be 10\% deeper than that produced by the rocky core occultation alone. It is worth noting that the effective altitude of an exoplanet atmosphere, $h$, will represent the main uncertainty in the calculations. For Earth, at specific wavelengths, the 40 $km$ altitude atmosphere is a conservative estimate, as atmospheric features that exist a few hundred $km$ high in the Earth's atmosphere (i.e. Ca) and in the lower troposphere, have both been detected in its transmission spectrum (Pall\'e et al, 2009).

\begin{figure*}[h]
\epsscale{1.0}  \plotone{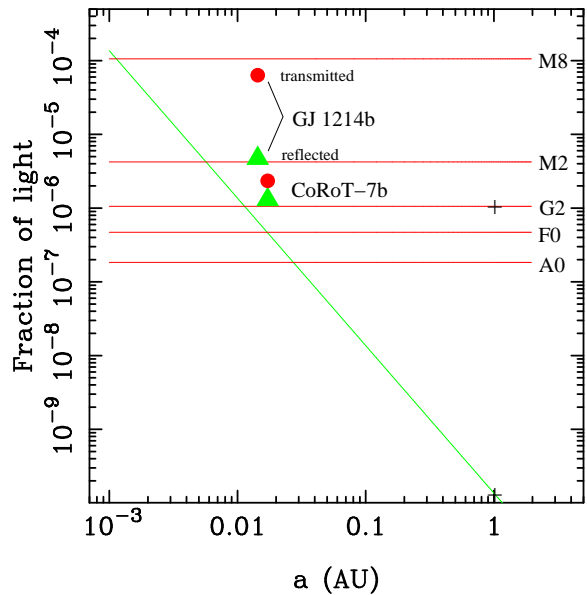} \caption{A direct comparison of the light reflected and transmitted by an earth-like planet. The green line indicates the amount of starlight reflected by the planet (normalized by the stellar flux using eq.~\ref{eq2}), which depends mainly on its size, its albedo assumed here constant to 0.3, and the distance to the star. The horizontal lines represent the amount of stellar light that crosses the planet atmospheric ring (supposed here of 40 km height), for different stellar types. The transmitted flux depends on the size of the planet atmospheric ring and the size of the star. The calculations neglect the intrinsic emission from the planet because, in the visible an near-IR wavelength range, it is negligible compared to the reflected/transmitted flux. In the mid- and far-infrared, this would no longer be the case (Spiegel et al, 2010). For illustration purposes the position in the diagram occupied in transmission and in reflection by the two transiting super-earths discovered so far, CoRoT-exo7b and GJ1214b, are shown. Note that the comparison of these two planets' positions to the modeled lines is not direct as the latter assume a 1 $R_{e}$ planet. The two crosses indicate the Earth's positions (reflected and transmitted light) in the diagram.}
\label{fig2}
\end{figure*}

The comparison between reflected and transmitted stellar fluxes by earth-like planets around a few types of stars, using the above formulation, is illustrated in Figure~\ref{fig2}. In view of the figure, it is possible that for some star-planet configuration, in particular, very close-in planets, the chances of detection might be larger in reflected light, but in general the opposite is true. For a copy of the Earth around a solar-type star (1 AU), transmission is always the most favorable configuration. More importantly, the flux contrast of transmission spectroscopy dominates in the habitable zones, where we expect to find extrasolar planets with atmospheric structure and composition close to the Earth's, and which depend on the spectral type of the parent star. This renders the transmission observation as the most technologically promising technique for the study of exoEarth twins at optical and near-infrared wavelengths.

\section{Exoplanet atmospheric characterization from ground and space.}

In this section we use the empirical stellar and planetary spectra, which we combine to simulate future ground and space-based observations of transiting exoplanets. Once a transiting planet is found, to retrieve the planetary spectra, one needs to observe individual spectra of the star+planet system {\it in} and {\it out of} transit. For the most favorable Hot Jupiter planets, and with current instruments, only a few transits are necessary.  For the small rocky planets, however,  due to the faintness of the planetary signal and the finite duration of the transit (hours), in practice this translates to having to observe several transits. These spectra will have typical exposure times of several minutes and will need to be combined in {\it in} and {\it out of} transit pairs to get their ratio spectrum. In turn, the resulting ratio spectra will need to be co-added for several transits until a sufficient signal to noise ratio (SNR) is reached.

M stars are particularly interesting planet hosts. Because of their small radii, low-mass planet companions can be detected more easily than around other stellar types (planet/star contrast is smaller). Moreover the probability of observing a planetary transit increases as the orbital radius of the planet decreases (Brown, 2001). Note that the habitable zone of M stars lies closer to the star than for other stellar types.

We have taken the transmission spectrum of the Earth measured by Pall\'e et al. (2009) and combined it with several stellar spectra  to simulate the repeated observations of a transiting star+planet system. In particular, we have used the models from Allard et al (2001), for three effective temperatures, $T_{eff}$, namely 5000, 3100 and 2000 $K$. All three models have $\it log(g)=5.0 (cgs)$ and solar metallicity, and are condensed model atmosphere where the layers containing dust particles are sitting below the photosphere. A temperature of 5000 K corresponds to a G type star for which we take the radius to be that of the Sun. A temperature of 3100 K corresponds to an M4 type star, with a radius of 0.2 solar radius. A temperature of 2000 K corresponds to a surface temperature transition between late-M and early L types (Golimowski et al. 2004), with a radius taken here to be 0.1 solar radius. Considering the effective temperature--spectral type calibrations provided by Stephens et al. (2009), Cushing et al. (2008), Vrba et al. (2004), Dahn et al. (2002), and Leggett et al. (2001 and 2002), field sources with spectral type L2 likely have an atmosphere of $T_{\rm eff}$ = 2000 K.

In order to simulate the resulting ratio spectrum from two individual spectral pairs (in and out of transit), $\frac{S_i}{S_o}$, we use the following expression:

\begin{equation}
\frac{S_i}{S_o}  = \frac{  (S_{*} - S_{*} \frac{A_{p+a}}{A_{*}} +  S_{*}\frac{A_{a}}{A_{*}} T_p ) + N_i     }{S_{*} + N_o }
\end{equation}

\noindent where, $S_{*}$ is the stellar spectrum, $T_p$ is the earth's transmission spectrum, and  $A_{*}$, $A_{p+a}$ and $A_{a}$ are the projected areas of the star, the planet with its atmosphere, and the planetary atmosphere alone, respectively. The terms $N_i$ and $N_o$ refer to random Poisson noise added individually to the denominator and the numerator in the equation. Here, the terms $S_{*}$, $T_p$, $N_i$ and $N_o$ are $\lambda$-dependent.

Apart from detection noise, there are a large number of factors that can induce variability in the star+planet spectrum, making it difficult in practice to isolate the planetary signal. The stellar spectrum can be affected by stellar activity and jitter, for example (Desort et al., 2007). In this study we have considered the star as ``quiet" over timescales of the duration of a transit (a few hours) and assumed that the effect of different activity levels from one transit to the next is removed due to the differential nature of the measurements. The Earth's atmosphere however is also an important source of uncertainty due to molecular absorption features. We discuss this in the next subsection.

Instead of assessing the capabilities of different instrumentation to reach the goal of atmospheric characterization (e.g. Belu et al. 2010), we have chosen to express the requirements in terms of the SNR necessary in individual observations, which can then be calculated specifically for any designed instrument. For stellar types G we compute the SNR value at the peak of the stellar flux, in the interval 0.5 to 0.6 $\mu m$. Thus, the rest of the spectrum has a lower, in specific region of the spectrum much lower, SNR. For spectral types M, the SNR is calculated over the wavelength interval 1.2 to 1.3 $\mu m$. The noise terms ($N_i$ and $N_o$) have been chosen so that both the {\it in-} and {\it out-of-} transit spectra have a SNR of 100, 1,000 or 10,000 at 1.2 $\mu m$. The ratio spectra analysis extends over the wavelength interval from 0.4 to 2.4 $\mu m$, for which the empirical Earth transmission spectrum is available. We have run the majority of our simulations with a spectral resolution of $R \sim 1,000$, to match that of the empirical Earth's spectrum.

\begin{figure*}[H]
\epsscale{2.0}  \plotone{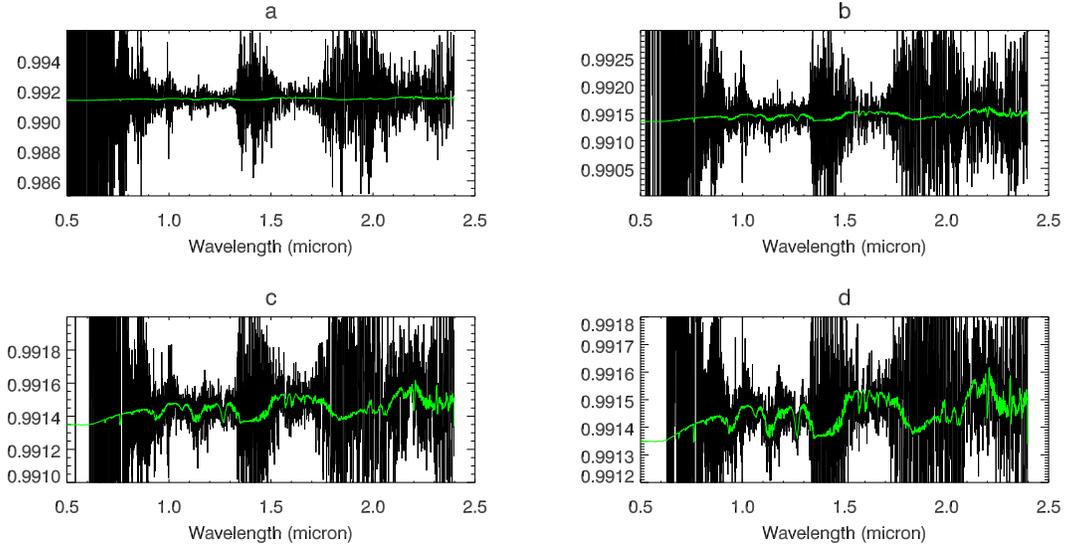}
\caption{Model simulations of the detectability of the atmospheric spectral features of an Earth-like planet around a $T_{\rm eff}\sim 2000 K$ star. In all panels the green solid lines is the real transmission spectrum of the Earth. The black solid line is the average of 20 (a), 100 (b), 500 (c), and 1000 (d) individual ratio spectra. Each individual spectrum had a SNR of 1000 at 1.2$\mu m$. As more and more spectra are combined, the planetary features become more evident as the noise is reduced.}
\label{fig3}
\end{figure*}

\begin{figure*}[H]
\epsscale{2.0}  \plotone{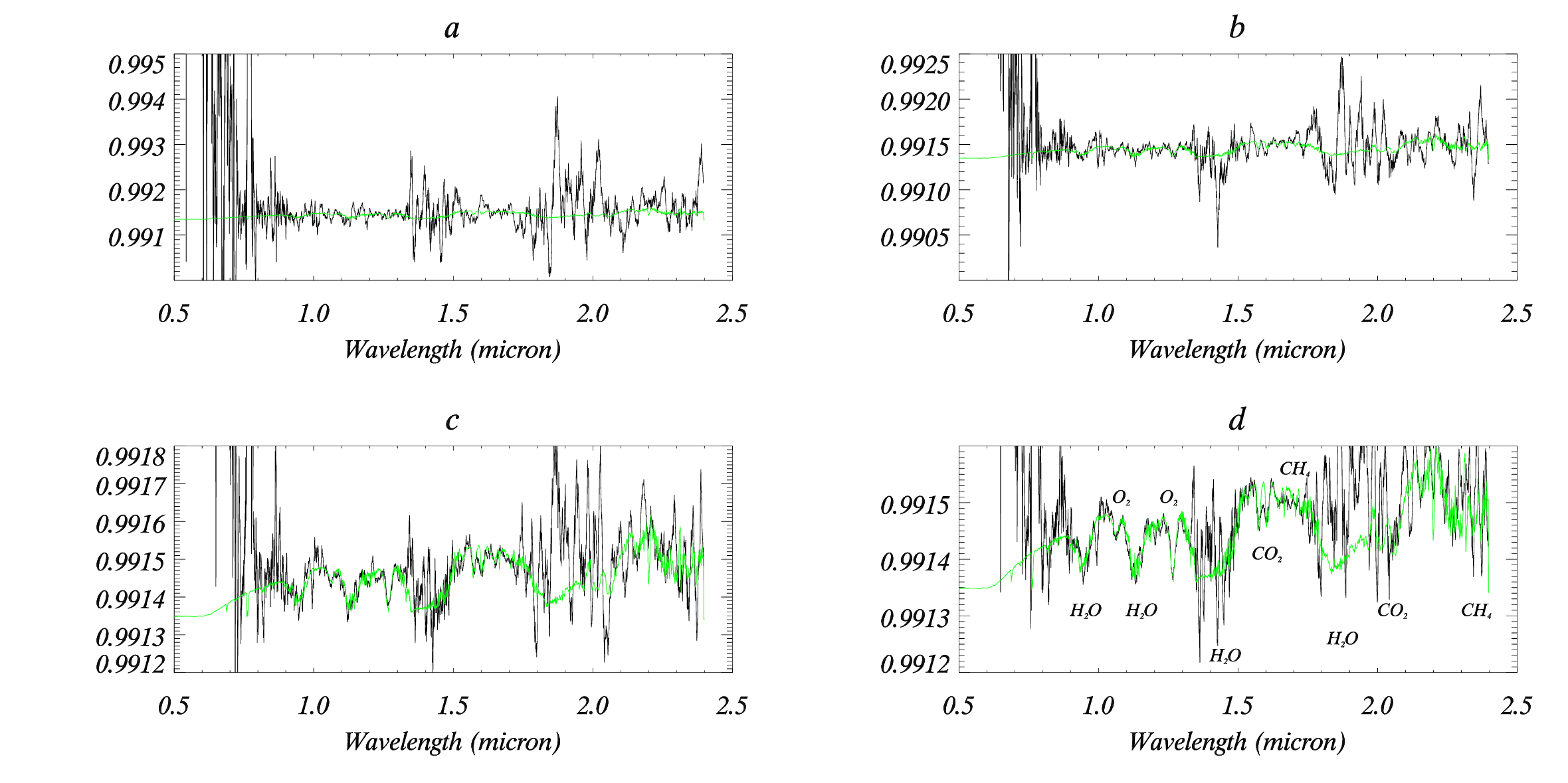}
\caption{Same as Figure~\ref{fig3}, but the resulting final co-added ratio spectra (black line) have been smoothed using a 10-point running mean.
}\label{fig4}
\end{figure*}

\begin{figure*}[H]
\epsscale{2.0}  \plotone{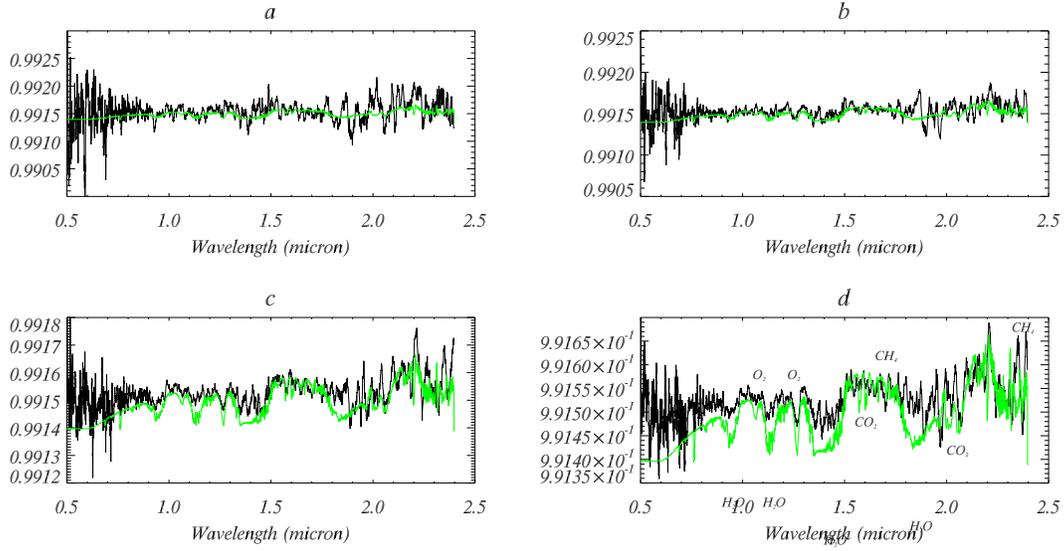}
\caption{Co-added ratio spectra (black line) smoothed using a 10-point running mean as in Figure~\ref{fig4}. Here however a $T_{\rm eff} \sim$ 3100\,K star ($\sim$M4) is used as the parent, and a 2 Earth's radius planet is assumed.
}\label{fig5}
\end{figure*}

We have generated thousands of these noisy spectra and co-added them to retrieve a cleaner spectrum. In Figure~\ref{fig3}, we illustrate how using a large number of individual noisy spectra progressively increases the SNR of the resulting combined spectrum. In the figure, the combination of 20, 100, 500, and 1000 individual spectra are plotted in black. The same Earth transmission reference spectrum is plotted in all panels in green. The results are given for an earth-like planet around an star with $T_{\rm eff} \sim2000 K$, where the initial simulated spectra have SNR at 1.2$\mu m$ of 1,000. It is seen in Figures~\ref{fig3} to ~\ref{fig5} that the major atmospheric features of the planetary atmosphere are discernible by combining around 100 individual pairs, and  are well constrained with 500. The larger amplitude in the noise levels shorter than 0.9 $\mu m$ and in the water bands regions near 1.5 and 2.0 $\mu m$ is due to absorption bands in the star's atmosphere, which reduce the stellar flux, thus decreasing the SNR of the spectrum. The quality of the spectra shown in the bottom panels of Figure~\ref{fig4} are similar to those of the hot Jupiter transmission spectra presently retrieved by HST or Spitzer (Swain et al , 2008; Tinetti et el, 2009; for example).

Obviously, the SNR necessary to detect atmospheric features differs for each atmospheric species, depending on the strength of their respective absorption features. Spectral regions where the parent star is devoid of molecular bands are more favorable for the detection of planetary features. In Table~\ref{table1}, we summarize the detectability of each spectral feature, depending on stellar spectral type, the planet's atmospheric thickness, and the number of co-added spectra. In the table, the sigma confidence level is given for the major atmospheric features of $H_2O$, $O_2$, $CO_2$ and $CH_4$, following Pall\'e et al (2009). Only confidence levels above $2\sigma$ are given in the table.

For late-M and L stars, in order to discriminate the planetary spectra, it is necessary to achieve a SNR of $10^{4}$ or larger in the combined spectrum (see Figure~\ref{fig3}). Retrieval of planetary features below 0.8-0.9$\mu m$ is too challenging even for an extremely large telescope due to the reduced stellar flux at these wavelengths. A SNR or $10^{4}$ across the whole spectrum would be necessary to extract the planetary spectrum. The stellar water features near 1.4 and 1.9 $\mu m$ also make it difficult to detect any planetary features. Thus, the most favorable wavelength range for the planet characterization is the spectral window from 0.9 to 1.3  $\mu m$. Within this range, the water vapor features at 1.13 $\mu m$ and the oxygen features at 1.06 and 1.26 $\mu m$ are the easiest transitions to identify. Methane at 2.26 $\mu m$ and carbon dioxide at 1.6 $\mu m$ can also be identified but they require slightly larger SNR, and therefore more integration time.

To put this result in context, we can take as an example the projected 42-meter E-ELT telescope. The typical object for this kind of characterization campaign would be an M star with magnitude in the Johnson's $V$ band of 12 or brighter. On the other hand, the integration time of the observations of a transiting planet has to be in the order of minutes, as the transit of a rocky planet around an M star only last for typically about 0.5 to 2 hours. Integration times at shorter time scales (seconds) would most probably be photon-starved. The online E-ELT Exposure Time Calculator (www.eso.org) predicts a SNR of about 1,000 for an on-source integration time of 3 minutes, for single exposures of a star with $m_v =12$ and $\frac{\lambda}{\Delta\lambda} =1000$. Thus, about 500 spectra (roughly 25 hours of integration time) will need to be combined to achieve the final desired SNR necessary to detect enough spectral features (water vapor, oxygen and methane) as to characterize the planet as inhabited (if earth-like). Including both the {\it in-transit} and {\it out-of-transit} spectra, that makes a total of 50-100 hours of observations, including overheads. However, adding the uncertainty introduced by  the telluric contribution always present in ground-based observations can increase the difficulty of the experiment (see next section).

%The orbital period of late-type M dwarfs ranges from a few hours to a few days, with possible planetary transits lasting on the order of hours. This means that transit events could occur almost daily, and the necessary SNR could be reached within 1-2 years of observations.

For mid and early-type M stars the SNR necessary to retrieve the planetary spectrum rises to about $10^{5}$, and require longer integration times. In Table~\ref{table2} we have computed the {\it in-transit} integration time required to observe the presence of water, water+oxygen, and water+ oxygen + either methane or carbon dioxide, for different stellar types and planet radii.

These results are valid for Earth twin planets, but planets exist in a large variety of sizes and evolutionary histories. In particular, we have already discovered several so-called super-earths, rocky planets larger than the Earth but still theoretically capable of supporting life (Tarter et al, 2007). If we consider a planet with a $2 R_E$, with a $40 km$ extended atmosphere, around a star with $T_{\rm eff} \sim 2000 K$, the prospects for detection increase quickly. Making the necessary assumption that atmospheric composition would be the same as that of the earth, we could detect with large confidence the simultaneous presence of $H_2O$, $CH_4$, $O_2$ and $CO_2$ in its atmosphere with about 20 spectra of $SRN = 1,000$ at 1,2 $\mu m$. This would imply the detection of life in its atmosphere, and could be accomplished in only a few hours of integration time.

\subsection{Ground-based Observations and telluric lines}

The results given in Table~\ref{table1}, however, only hold true for space-based telescopes. One of the major challenges for ground-based telescopes will be to accurately account for the telluric spectral features of the Earth's atmosphere. Each individual spectrum will need to be ``decontaminated" of the local atmospheric signatures to clearly retrieve the target's spectrum.

To estimate the effect on feature detectability, we have included in our models the effects of telluric lines. Ideally, we should combine a high-resolution transmission spectrum of the Earth with a high-resolution telluric spectrum, the two spectra being Doppler-shifted from each other as dictated by the target drift velocity. Subsequently, the combined spectrum should be convolved to the spectral resolution and PSF appropriate to the observing instrument (Bailey, 2007). However, this procedure is not feasible if we are to use the empirical Earth's transmission spectrum measured by Pall\'e et al. (2009), which has a moderate spectral resolution and no Doppler shift. Here, we have decided to use for the telluric spectrum a modeled spectrum at the spectral resolution of the measurements. In order to accomplish this, we have convolved each of our individual spectra with a synthetic telluric spectrum, before adding the noise. To simulate realistic conditions, the telluric spectra were different for the {\it in-} and {\it out-of-transit} spectra for each individual ratio. Under such assumption, Eq. (4) can be rewritten as:

\begin{equation}
\frac{S_i}{S_o}  = \frac{  ( (S_{*} - S_{*} \frac{A_{p+a}}{A_{*}} +  S_{*}\frac{A_{a}}{A_{*}} T_p )  S_ti)+ N_i     }{(S_{*} S_to) + N_o }
\end{equation}
where, $S_ti$ and $S_to$ are the in- and out-of-transit telluric spectra of the local atmosphere (on top of the telescope), and are equal to 1.0 at all $\lambda$s for a space-based telescope.

The telluric spectra were calculated line-by-line in the spectral direction by evaluating the optical thickness at the zenith (Garcia-Mu\~noz et al,  {\it in preparation}), using the latest version of HITRAN (Rothman et al, 2009). The model assumes the US1962 standard atmosphere, for all of gas constituents but water. For the latter molecule, we randomly varied its abundance by a factor from 1/2 to 2 relative to the standard profile, although such variability range is probably exceedingly large (Garcia-Lorenzo et al, 2010). Over the course of a few hours the changes in atmospheric concentration of $O_2$, $CH_4$ and $CO_2$ are assumed to be negligible.

\begin{figure*}[H]
\epsscale{2.0}  \plotone{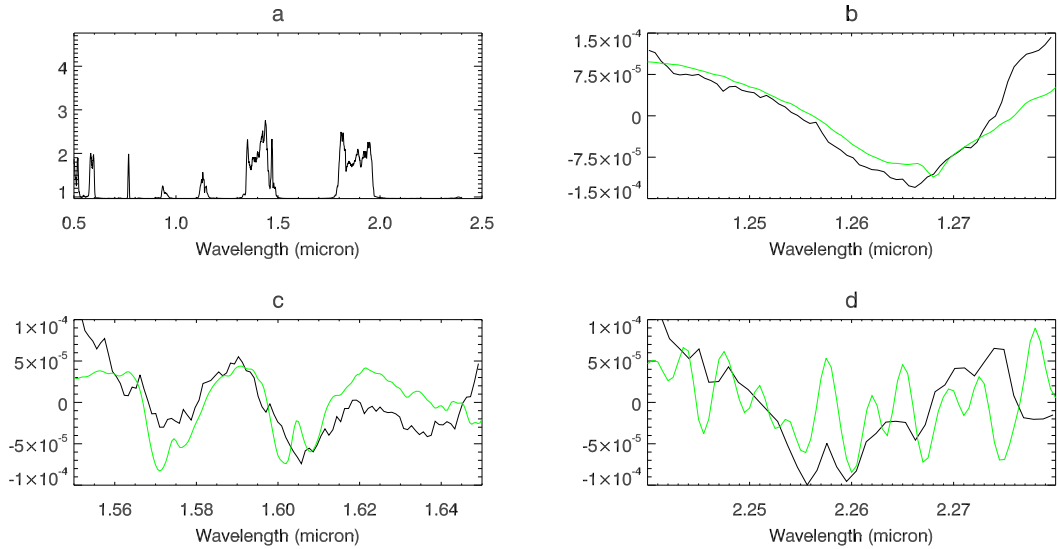}
\caption{Model simulations of the detectability of the atmospheric spectral features of a super-Earth-like planet around $T_{\rm eff}\sim 2000 K$ star, by co-adding 500 individual ratio spectra. Data have been smoothed using the same running mean as in Figure~\ref{fig4}. The four subpanels correspond to the full range of modeled wavelengths (a), and a zoom in on the region around the $O_2$@$1.26 \mu m$ (b), $CO_2$@$1.6 \mu m$ (c), $CH_4$@$2.26 \mu m$ (d) molecular absorption features. Over the full spectral range, only the strong features of the local telluric water vapor, unsuccessfully removed from the data are visible. In the other three subpanels, the vertical scale indicates the spectral absorption anomalies over each displayed spectral range. The green line is the earth's transmission spectrum normalized to the vertical scale.}
\label{fig6}
\end{figure*}

The major effect of adding the telluric spectra is that detecting $H_{2}O$ vapor in the atmosphere of rocky planets is no longer possible. The telluric variability on Earth completely prevents our model from reaching sufficient SNR even with thousands of spectra. Nevertheless, the rest of the atmospheric constituents are still detectable with the same integration times given in Tables~\ref{table1} and ~\ref{table2}. This is because the molecular transitions of these features occur far from those of water vapor. In Figure~\ref{fig6} we plot the results of co-adding 500 ratio spectra observed through the Earth's atmosphere we have defined. The figure zooms on three different spectral features occurring far from any water vapor transitions.

Hopefully, the Earth's atmosphere, however, will not remain an unsurmountable impediment for long. Although it was always considered technically too challenging, the evaluation and careful removal of systematic errors have pushed the detectability limits, and ground based characterization at near-infrared wavelengths is already a reality (Sing and L{\'o}pez-Morales, 2009; Swain, 2010). In fact, there are several data analysis strategies that when applied can push the results of ground-based observatories toward those in Table~\ref{table1}. In particular, each individual spectral observation can be a priori decontaminated using synthetic models, or when the instrumental configuration allows for it, using calibration stars observed simultaneously. Moreover, because of their different nature, there are some spectral features that appear in the Earth's globally-averaged transmission spectrum but not in the telluric absorption spectrum on top of the telescope. For example the absorption bands of atmospheric dimers($O_2-O_2$ and $O_2-N_2$; Smith and Newnham, 2000), are weak features in the local atmosphere telluric spectrum, because the path of light through the atmosphere is short (Pall\'e et al, 2009). These spectral features, in combination with standard calibration techniques, can be used to improve the data analysis and make sure that the final retrieved spectrum is clear of telluric contamination.

Finally, we must note that we are conducting our simulations at spectral resolution of $R \sim 1,000$. Thus, we cannot take advantage of the Doppler displacement of the star+planet absorption features to separate them from the telluric features. Future highly-stable high-resolution spectrographs mounted on large aperture telescopes will surely take advantage of this phenomenon (e.g., Snellen et al. 2010) to improve the detectability limits set up in Tables~\ref{table1} and ~\ref{table2}. Resolving powers of the order of $60,000$ are enough to distinguish Doppler shifts of a few $km/s$.

\section{Conclusions}

In the very near future, the characterization of extrasolar planet atmospheres through transit spectroscopy will enter the domain of super-Earth and earth-like planets. The goal will be to detect atmospheric molecules such as $O_3$, $O_2$, $H_2O$, $CO_2$, and $CH_4$ that, if seen in combination, cannot only provide information about the composition and evolution of the planet, but also hint at the presence of life.
Here, we have used a semi-empirical approach to explore the feasibility of such characterization, with current and planned telescopes facilities.
We have shown how transmission spectroscopy during primary transits is a much more favorable approach (at least in the visible and near-IR ranges) for exoplanet characterization than secondary transit (except for very close-in, ``hot", uninhabitable, rocky planets).

Our simulations using an empirical terrestrial transmission spectrum, and observed stellar spectra for a variety of stellar types, indicate that the new generation of extremely large telescopes, such as the proposed 42-meter E-ELT, could be capable of retrieving the transmission spectrum of an Earth-like planet around stars with $T_{\rm eff} \le \sim 3100$\,K.
 For G type stars, however, the required SNR will still be beyond technical reach.

For a twin of Earth around an M4 star ($T_{\rm eff} \sim 3100 K$), for example, the spectral features of $H_2O$, $CH_4$, $CO_2$, and $O_2$ in the wavelength range between 0.9 to 2.4 $\mu m$ can simultaneously be detected within a hundred hours of observing time. The detection time can be reduced to a few hours for stars with $T_{\rm eff} \le \sim 2000$\,K or for a super-Earth planet with twice the Earth's radius. This would constitute a proof for the existence of life on that planet.

The major remaining issue toward characterization will be the effects of the Earth's atmosphere, as large aperture ELT telescopes will surely be ground-based. Our simulations indicate that including the telluric spectrum of the atmosphere into the models prevents reaching sufficient SNR in all wavelength ranges where water transitions are present, being water vapor the major atmospheric variable. Nevertheless, in wavelength ranges where water vapor variability is not present the detection of $O_2$, $CH_4$, and $CO_2$ is still possible.

\acknowledgments

Research by E. Pall\'e was supported by a `Ramon y Cajal' fellowship. Partial financial support for M. Zapatero-Osorio comes from Spanish project 200950I010.

%\begin{landscape} %+++++++++++

\begin{table*}
\caption{Detectability of several spectral bands of water vapor, methane, oxygen and carbon dioxide in the combined ratio spectrum of an earth-like planet transiting around an M star. Detectability sigma values are given for each band as a function of stellar type, planet radius, and thickness of the planetary atmosphere. Only values above the 2-$\sigma$ level are given. In this table the conclusions apply to a telescope in space, i.e., the influence of telluric absorption from the local atmosphere is not considered. The central wavelength of each band (in $\mu m$) is given below the molecule name. } \vspace{2mm}
\begin{tabular}{|l|ccc|cc|cc|cccc|}
\hline
Num spectra &  $H_2O$ & $H_2O$ & $H_2O$ & $CH_4$ & $CH_4$ &  $O_2$ & $O_2$ & $CO_2$ & $CO_2$ & $CO_2$ & $CO_2$    \\
            &  1.13   & 1.35   & 1.85   & 1.64   & 2.26   &  1.06  & 1.26  & 1.57 & 1.60 & 2.01 & 2.06    \\
\hline
  & \multicolumn{10}{c}{$T_{\rm eff} \sim2000 K$  $R_{p}= 1 R_{e}$ $h= 40 km$  } \\
\hline
20 &           - &      - &      - &            - &      - &          -&       - &      - &       - &       - &      - \\
100 &             - &       - &       - &           - &      - &          - &       - &       - &       - &      - &      - \\
500 &            - &       - &       3.78 &           2.07 &       - &         - &       2.66 &       - &       - &       - &       - \\
1000 &           - &      - &       3.69 &            2.59 &      - &        -&       3.52 &      - &       2.32 &       - &     - \\
10000 &            6.70 &       4.91 &      10.76 &           2.39 &       - &         - &       7.49 &       - &       2.56 &      - &       - \\
\hline
  & \multicolumn{10}{c}{ $T_{\rm eff} \sim2000 K$   $R_{p}= 1 R_{e}$ $h= 100 km$ } \\
\hline
20 &            - &    - &     - &         - &       - &      - &       - &       - &      - &      - &      - \\
100 &            - &       - &       3.18 &            -&      - &         - &       2.45 &    - &       - &       - &       - \\
500 &              2.28 &       3.37 &       7.13 &           -&       2.99 &        - &       7.29 &       4.28 &       4.28 &       -&      - \\
1000 &          2.95 &       5.75 &      10.70 &            -&       4.93 &          2.81 &       6.87 &       4.86 &       4.07 &      - &       2.16 \\
10000 &    9.72 &      11.69 &      22.27 &       4.24 &       7.12 &        5.40 &      15.75 &       2.98 &       3.22 &       7.33 &       8.64 \\
\hline
& \multicolumn{10}{c}{$T_{\rm eff} \sim2000 K$   $R_{p}= 2 R_{e}$ $h= 100 km$ } \\
\hline
20 &       - &     - &       4.27 &          - &      - &       - &       2.53 &       2.49 &       - &     - &     - \\
100 &        - &       2.03 &       4.71 &    - &     - &     - &       8.23 &       4.65 &       3.87 &    - &     - \\
500 &       - &       9.82 &      11.19 &          - &      - &         3.79 &       6.49 &       4.42 &       3.85 &       -&       - \\
1000 &   - &      10.10 &      12.12 &      3.82 &       2.66 &          4.17 &      11.26 &       5.11 &       4.18 &       2.05 &       4.50 \\
10000 &   7.58 &      14.47 &      45.66 &   5.33 &       6.89 &        6.67 &      24.84 &       7.08 &       6.55 &       3.00 &      11.44 \\
\hline
  & \multicolumn{10}{c}{$T_{\rm eff} \sim3100 K$   $R_{p}= 1 R_{e}$ $h= 100 km$ } \\
\hline
20 &          -&      - &       -&           - &      - &            -&       - &      - &      - &     - &      - \\
 100 &        -&       2.34 &      - &          - &      -&    -&      - &      - &      - &      - &       - \\
 500 &         - &       2.42 &       3.68 &     - &   - &            - &     - &       - &    - &    - &       - \\
 1000 &          - &       1.36 &       9.87 &       - &      - &   - &       - &       3.79 &       5.17 &      - &      - \\
 10000 &          - &       7.24 &       4.08 &           3.12 &       4.54 &           2.50 &       3.43 &       2.82 &       6.33 &      - &       3.10 \\
\hline
\end{tabular}
\label{table1}
\end{table*}

%\clearpage

%\end{landscape} %++

\begin{table*}
\caption{Amount of in-transit integration time necessary to achieve the detection of i) water vapor; ii) water vapor and oxygen; and iii) water vapor, oxygen and methane and/or carbon dioxide in the transmission spectrum of an extrasolar earth-like planet, with the E-ELT. The format of the table responds to the fact that in most cases water vapor is the most easily detectable feature, followed by oxygen and then carbon dioxide and methane. In other cases the SNR necessary for the detection is reached simultaneously (note that here we consider only the time scales corresponding to the time needed to integrated 5, 20, 100, 500, 1,000 and 10,000 individual spectra). A detection confidence level of $2\sigma$ is set as the lower limit. The amount of time is expressed in hours, and the total necessary telescope time (not including overheads) would be twice this amount to acquire the necessary out-of-transit observations. Note that the times in the last column on the table refer to the detection of the triple fingerprint, equaling our best indication of the planet being inhabited.} \vspace{2mm}
\begin{tabular}{|l|ll|c|c|c|}
\hline
Star $T_{\rm eff}$ & \multicolumn{2}{c}{Planet size}  & $H_2O$ & $H_2O$ + $O_2$  & $H_2O$ + $O_2$ + ($CH_4$/$CO_2$)    \\
\hline
 & $R_{p}$ & $ h [$km$]$   & \multicolumn{3}{c}{ Time [$hours$]} \\
\hline
\hline
$2000 K$   & $R_{e}$    &   40  &    25    &   25     &   25   \\
$2000 K$  & $R_{e}$    &  100  &     5    &    5     &    5   \\
$2000 K$  & 2*$R_{e}$  &   40  &     1    &    5     &    5   \\
$2000 K$  & 2*$R_{e}$  &  100  &     1    &    1     &    5   \\
\hline
$3100 K$  & $R_{e}$    &   40  &   170    &  170     &  500   \\
$3100 K$  & $R_{e}$    &  100  &     5    &   50     &  500   \\
$3100 K$  & 2*$R_{e}$  &  100  &     5    &    5     &    5   \\
\hline
$5800 K$  & $R_{e}$    &  100  &     $>$500    &    $>$500     & $>$500   \\
$5800 K$  & 2*$R_{e}$  &  100  &     $>$500    &    $>$500     &  $>$500   \\
\hline
\end{tabular}
\label{table2}
\end{table*}

\clearpage

\end{document}